\documentclass[letterpaper, 10 pt]{ieeeconf}

\usepackage{amsfonts,amsmath,amssymb} 

\usepackage{psfrag,color}
\usepackage{enumerate,cite,latexsym,graphicx}
\newtheorem{theorem}{Theorem}
\newtheorem{lemma}{Lemma}

\newtheorem{definition}{Definition}

\title{\LARGE \bf A Systems Theory Approach to the Synthesis of Minimum Noise Phase-Insensitive Quantum Amplifiers}

\author{Ian R.~Petersen,  Matthew R.~James, Valery Ugrinovskii and Naoki Yamamoto%
\thanks{This work was supported by the Air Force Office of Scientific
Research (AFOSR), under agreement number FA2386-16-1-4065.}%
\thanks{Ian R. Petersen and Matthew R. James are with the Research School of  Engineering, 
        The Australian National University, Canberra ACT 2601, Australia.
         {\tt\small i.r.petersen@gmail.com, matthew.james@anu.edu.au} }
\thanks{Valery Ugrinovskii is with the School of  Engineering and Information Technology, 
        University of New South Wales at the Australian Defence Force Academy, Canberra ACT 2600, Australia.
{\tt\small v.ugrinovskii@gmail.com}} 
\thanks{Naoki Yamamoto is with the Department of Applied Physics and Physico-Informatics,
Keio University, Yokohama 223-8522, Japan. 
{\tt\small yamamoto@appi.keio.ac.jp} }
}%

\begin{document}

\maketitle
\thispagestyle{empty}
\pagestyle{empty}

\begin{abstract}
We present a systems theory approach to the proof of a result bounding the required level of added quantum noise in a phase-insensitive quantum amplifier. We also present a  synthesis procedure for constructing a quantum optical  phase-insensitive quantum amplifier which adds the minimum level of quantum noise and achieves a required gain and bandwidth. This synthesis procedure is based on a singularly perturbed quantum system and leads to an amplifier involving two squeezers and two beamsplitters. 
\end{abstract}

\section{Introduction} \label{sec:intro}
In the theory of quantum linear systems \cite{GJN10,ZJ11,ShP5,PET10B,NY17}, quantum optical signals always have two quadratures. These quadratures can be represented either by annihilation and creation operators, or position and momentum operators; e.g., see \cite{PET10B,NY17}. The relative size of the two quadratures in a quantum optical signal determines the optical phase of the signal. In designing an amplifier for a quantum optical signal, is often desired to preserve the optical phase of the amplified signal. Such quantum amplifiers are referred to as phase-insensitive amplifiers or  phase-preserving amplifiers. In the paper, \cite{CAV82} (see also \cite{HM62}), Caves recognized the importance of phase-insensitive amplifiers and showed that Heisenberg's uncertainty principle implies that any phase-insensitive amplifier must also introduce an amount of quantum noise which is related to the level of amplification required. The use of phase-insensitive quantum amplifiers plays a key role in areas of quantum technology such as quantum communication and weak signal detection; e.g., see \cite{BSM+7,CDGMS10,CWA+5,MC14,HKL+3,NY17}. Phase-insensitive quantum amplifiers can be implemented using non-degenerate optical parametric amplifiers (NOPAs) \cite{YAM16}; squeezers, beamsplitters and measurement feedforward \cite{YMFF11}; or using feedback optical systems \cite{NY17}. 

In this paper, we re-derive the noise bound of \cite{CAV82} for phase-insensitive quantum amplifiers using the quantum linear systems notion of physical realizability and in particular the physical realizability of a transfer function matrix; e.g., see \cite{GGY08,GJN10,ZJ11,ShP5,PET10B,KhP1}. We also present a systematic procedure for synthesizing a quantum optical phase-insensitive quantum amplifier with a given gain and bandwidth, using a pair of degenerate optical parametric amplifiers (squeezers) and a pair of beamsplitters. This approach is based on the singular perturbation of quantum systems \cite{PET09A,VuP3a} to achieve the required DC gain and bandwidth. Compared with the NOPA approach such as described in \cite{NY17}, our approach uses squeezers for which it is typically easier to obtain a higher level of squeezing (and hence amplifier gain). Also, compared to the approach of \cite{YMFF11}, our approach does not require quantum measurement. In addition, compared to the feedback approach of \cite{YAM16,NY17}, our approach always achieves the minimum amount of required quantum noise and only requires a fixed level of squeezing for a given amplification. Our proposed quantum optical phase-insensitive quantum amplifier synthesis procedure may be useful in on-chip quantum optical technologies such as described in \cite{DLMGNL15}.

\noindent {\bf Notation:~}
 $I$ denotes the identity matrix, 
 $J :=
  \left[ \begin{array}{cc} I & 0\\[-1mm]
    0 & -I  \end{array} \right] $.
 For a  matrix $X$ of operators, $X^T$ and
 $X^\#$ respectively denote the matrices
 (of operators) obtained by
  taking transpose and component-wise adjoint.
  $X^\dagger := (X^\#)^T$.
Also, if $X$ is a complex matrix, then
 $X^T$ denotes the usual transpose and
   $X^\#$ denotes the matrix obtained by component-wise complex 
   conjugation.
 For a single operator (resp. complex scalar)
   $g$,
 we  use $g^*$ to denote its
  adjoint (resp. complex conjugate).
   If $x,y$ are column vectors
  (of same length) of operators, then we define the
   commutator
 $[x,y^T] :=xy^T - (yx^T)^T$. Consequently,
   $[x,y^\dagger] = [x,(y^\#)^T]=
   xy^\dagger - (y^\# x^T)^T$.

   Matrices of the form
   $\left[ \begin{array}{cc}
 R_1 & R_2 \\ R_2^\# & R_1^\#
     \end{array}  \right]$
 are denoted by $\Delta(R_1,R_2)$; see also \cite{GJN10,ZJ11}.

\section{Linear  Quantum Systems} \label{sec:probstat}
We consider a class of linear  quantum  systems described by the quantum stochastic differential equations (QSDEs), (e.g., see
\cite{GJN10,PET10B,ShP5}):
\begin{eqnarray}
\label{sys} \left[\begin{array}{l} d a(t)\\d
a(t)^\#\end{array}\right] &=&  A\left[\begin{array}{l} 
a(t)\\ a(t)^\#\end{array}\right]dt +  B
\left[\begin{array}{l} du(t)
\\ du(t)^{\#} \end{array}\right];  \nonumber \\
\left[\begin{array}{l} dy(t)
\\ dy(t)^{\#} \end{array}\right] &=&
 C \left[\begin{array}{l}  a(t)\\
a(t)^\#\end{array}\right]dt +  D \left[\begin{array}{l}
du(t)
\\ du(t)^{\#} \end{array}\right],\nonumber \\
\end{eqnarray}
where
\begin{eqnarray}
\label{FGHKform}  A =\Delta( A_1,  A_2),
 &&
 B = \Delta( B_1,  B_2), \nonumber \\
 C = \Delta( C_1, C_2),
 &&  D = \Delta( D_1, D_2).
\end{eqnarray}
Here, $A_1 \in \mathbb{C}^{n \times n}$, $A_2 \in \mathbb{C}^{n \times n}$, 
$B_1 \in \mathbb{C}^{n\times m}$, $B_2 \in \mathbb{C}^{n\times m}$, 
$C_1 \in \mathbb{C}^{m \times n}$, $C_2 \in \mathbb{C}^{m \times n}$,
$D_1 \in \mathbb{C}^{m \times m}$ and $D_2 \in \mathbb{C}^{m \times m}$. 
Also, $ a(t) = \left[ {a_1 (t) 
\cdots a_n (t)} \right]^T$ is a vector of (linear combinations of) annihilation operators.
The vector $u$ represents the input signals and is assumed to admit the
decomposition:
\begin{equation}\nonumber
 du(t) = \beta _{u}(t)dt + d \tilde u(t)
\end{equation}
where $\tilde u(t)$ is the noise part of $u(t)$ and $\beta_{u}(t)$
is an adapted process (see \cite{BHJ07},
\cite{PAR92} and \cite{HP84}).
The noise $ u(t)$ is a vector of quantum noises.  The noise
processes can be represented as operators on an appropriate Fock 
space (for more details see \cite{VPB92} and
\cite{PAR92}).
The process $\beta_{u}(t)$ represents variables of other systems
which may be passed to the system (\ref{sys}) via an interaction. More
details concerning this class of quantum systems can be found in the
references \cite{JNP1,GJN10,PET10B,ShP5}).

\begin{definition} (See \cite{JNP1,ShP5,PET10B}.) 
\label{phys_real}
A complex linear quantum 
  system of the form (\ref{sys}), (\ref{FGHKform}) is 
  said to be {\em physically realizable} if there  exists a 
complex commutation matrix $\Theta= \Theta^\dagger$, a complex
Hamiltonian matrix $M = 
M^\dagger$, and a  coupling matrix $N$ such that 
\begin{equation}
\label{Psiform}
\Theta = TJT^\dagger
\end{equation}
where $T=\Delta(T_1,T_2)$ is non-singular, $M$ and $N$ are of the form
\begin{equation}
\label{tildeMN}
 M= \Delta( M_1, M_2), N= \Delta( N_1, N_2)
\end{equation}
and 
\begin{eqnarray}
\label{generalizedFGHK1}
 A &=& -\imath \Theta   M -\frac{1}{2} \Theta  N^\dagger J  N; \nonumber \\
 B &=& -\Theta   N^\dagger J; \nonumber \\
 C &=&  N; \nonumber \\
 D &=& I.
\end{eqnarray}
\end{definition} 
In this
definition, if the system (\ref{sys}) is physically realizable, then
the matrices $M$ and $N$ define a complex open harmonic
oscillator with  scattering matrix $S=I$, coupling operator vector
\[
L = \left[\begin{array}{cc} N_1 &  N_2 \end{array}\right]
\left[\begin{array}{c} a \\  a^\#\end{array}\right]
\]
and Hamiltonian operator 
\[
\mathcal{H} =
 \frac{1}{2}\left[\begin{array}{cc} a^\dagger &
       a^T\end{array}\right] M
\left[\begin{array}{c} a \\  a^\#\end{array}\right];
\]
e.g., see 
\cite{GZ00}, \cite{PAR92}, \cite{BHJ07},
\cite{GJ09}, \cite{JNP1} and \cite{EB05}.

\begin{theorem}[See \cite{NJP1,JNP1}.]
\label{T0} The linear quantum system (\ref{sys}), (\ref{FGHKform}) is
physically realizable if and only if there exists a complex matrix
$\Theta =\Theta^\dagger$  such that $\Theta$ is
of the form in (\ref{Psiform}), and
\begin{eqnarray}
\label{physreal1} && A\Theta + \Theta  A^\dagger + 
BJ B^\dagger = 0;
\nonumber \\
&& B =  -\Theta  C^\dagger J; \nonumber \\
&& D = I.
\end{eqnarray}
\end{theorem}

The complex transfer function matrix  corresponding to the system (\ref{sys}) is given by
\[
G(s) = C(sI-A)^{-1}B+D. 
  \]

\begin{definition}
\label{D4a}
A complex transfer function matrix $G(s)$ is said to be physically realizable if it is the transfer function of a physically realizable linear quantum system. 
\end{definition}

 A physically realizable transfer function matrix corresponds to a linear quantum system which satisfies the laws of quantum mechanics and can be implemented using physical components such as arising in quantum optics; e.g., see \cite{NJD09,PET08A,NUR10,NUR10A,NGP1,NY17,GP1}. 

\section{Problem Formulation}
\begin{figure}[htbp]
\begin{center}
\includegraphics[width=6cm]{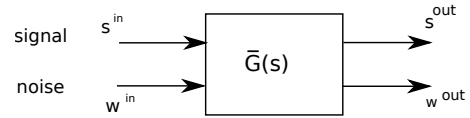}
\end{center}
\caption{Phase-insensitive quantum amplifier.}
\label{F1}
\end{figure}

A phase-insensitive quantum amplifier is a two-input two-output physically realizable quantum linear system with transfer function $\bar G(s)$ as illustrated in Figure \ref{F1}. In this diagram, the first input channel and the first output channel are the signal input and output channels respectively. Also, the second input channel and the second output channel are noise input and output channels. The noise output channel is not used in the operation of the amplifier but is included for consistency with the physical realizability theory for quantum linear systems; e.g., see \cite{JNP1,ShP5,KhP1}. As with any quantum linear system, each input and output channel consists of two quadratures; e.g., see \cite{ShP5,PET10B,NY17}. Hence, the transfer function matrix $\bar G(s)$ is a four-by-four transfer function matrix. In order to define a phase-insensitive quantum amplifier, a physically realizable transfer function matrix $\bar G(s)$ should satisfy certain gain and phase-insensitivity properties over a specified frequency range. These properties will be formally defined below. 

In this paper, we will formally consider the properties of $\bar G(s)$ to hold at DC; i.e., at $s= 0$. In addition, we will look at synthesizing amplifiers such that these properties also hold (approximately) out to some bandwidth frequency. However, it would be straightforward to extend out techniques so that the required phase-insensitive amplifier properties hold on any specified frequency interval. 

In the pioneering paper \cite{CAV82}, Caves showed that phase-insensitive quantum amplification can only be achieved at the expense of adding noise to the signal; see also \cite{HM62,YAM16}. We re-derive this result using the quantum linear systems theory notion of physical realizability and then give a systematic synthesis procedure for designing a physical phase-insensitive amplifier achieving a specified gain, which can be implemented using quantum optics. 

As in \cite{ShP5,PET10B}, we write the transfer function $\bar G(s)$ in ``doubled-up'' form, specifying both quadratures of each input and output channels as follows:
\begin{eqnarray}
\label{Gbar}
\left[\begin{array}{c} s^{out} \\ w^{out} \\ s^{out *} \\w^{out *}\end{array}\right] &=&
\bar G 
\left[\begin{array}{c} s^{in} \\ w^{in} \\ s^{in *} \\w^{in *}\end{array}\right] \nonumber \\
&=&
 \left[\begin{array}{cc}G & H \\
H^\# & G^\# \end{array}\right]
\left[\begin{array}{c} s^{in} \\ w^{in} \\ s^{in *} \\w^{in *}\end{array}\right].
\end{eqnarray}
Here and in the sequel, we will usually drop the dependence of transfer functions on the Laplace variable $s$ for  simplicity of notation. 
Using this notation, the two quadratures of the input signal are denoted by $\left[\begin{array}{c}s^{in} \\ s^{in *}\end{array}\right]$ and a similar notation applies to the other input and output signals. Furthermore, we write
\begin{equation}
\label{GH}
G=\left[\begin{array}{cc}g_{11} & g_{12} \\ g_{21} & g_{22} \end{array} \right], \quad 
H= \left[\begin{array}{cc}h_{11} & h_{12} \\ h_{21} & h_{22} \end{array} \right]. 
\end{equation}

We now present a result on the  physical realizability of a transfer function matrix; e.g., see  \cite{JNP1,GJN10,ShP5,KhP1}. 

\begin{lemma}[\cite{GJN10,ShP5,KhP1}]
\label{L0}
A transfer function matrix $\bar G(s)$ of the form (\ref{Gbar}) is physically realizable if and only if 
\[
\bar G^\sim(s)J\bar G(s) = J
\]
for all $s \in \mathbb{C}$ and
the matrix $\bar G(\infty)$ is of the form
$\bar G(\infty) = \left[\begin{array}{cc}S & 0\\ 0 & S^\#\end{array}\right]$ where $S^\dagger S = SS^\dagger = I$.  Here,  $\bar G^\sim(s)=\bar G(-s^*)^\dagger$.
\end{lemma}


Note that it follows from this result that any physically realizable transfer function $\bar G(s)$ will satisfy
\begin{equation}
\label{PR1}
\bar G(j\omega)^\dagger J\bar G(j\omega) = J
\end{equation}
for all $\omega$; see also \cite{GGY08,GJN10}.

\begin{definition}[see also \cite{CAV82}]
\label{D2}
A physically realizable transfer function matrix $\bar G(s)$ of the form (\ref{Gbar}), (\ref{GH}) is said to be {\em phase-insensitive} at frequency $\omega$ if 
\begin{equation}
\label{PI}
h_{11}(j\omega) = 0.
\end{equation} 
\end{definition}

We will be mostly concerned with the phase-insensitive property at DC and hence, we will usually drop the frequency specification. Also, we will be concerned with the corresponding {\em amplifier gain} squared amplitude 
$$g_{11}(j\omega)^*g_{11}(j\omega)$$
 and {\em noise} squared amplitude 
$$g_{12}(j\omega)^*g_{12}(j\omega)+h_{12}(j\omega)^*h_{12}(j\omega)$$
 at a given frequency $\omega$ (usually DC). 

\section{Main Results}

In this section, we first re-derive the main result of \cite{CAV82} in terms of the physical realizability notions given in the previous section. That is, we show that any physically realizable transfer function matrix which is phase-insensitive at a given frequency $\omega$ will have the property that the minimum possible value of the noise squared amplitude at that frequency is equal to the {\em amplifier gain} squared amplitude at that frequency  minus one. Also, for the case of $\omega =0$, (the DC case), we give a method for synthesizing a physically realizable transfer function which achieves this lower bound on the noise squared amplitude. Furthermore, this construction allows these properties to be (approximately) continued out to some arbitrary bandwidth. 

\begin{theorem}[see also \cite{CAV82}]
\label{T1}
At any frequency $\omega$, given a desired phase-insensitive quantum amplifier gain at that frequency $g_{11}$, then 
\begin{eqnarray}
  \label{noise_bound}
\lefteqn{\min \left[ g_{12}^*g_{12}+h_{12}^*h_{12} \right]} \nonumber \\
&=& g_{11}^*g_{11} - 1.
\end{eqnarray}
Here the minimum is taken over all transfer function matrices (\ref{Gbar}), (\ref{GH}) satisfying the physical realizability condition (\ref{PR1}), the phase-insensitivity condition (\ref{PI}) and with the given amplifier gain $g_{11}$. Furthermore, this minimum is achieved by the transfer function matrix defined by 
\begin{eqnarray}
  \label{optimal}
g_{12} &=& 0; g_{21}= \sqrt{\frac{g_{11}^*g_{11} -1}{g_{11}^*g_{11}}}; g_{22}=\sqrt{1+g_{11}^*g_{11}}; \nonumber \\
h_{11}&=&0;h_{12} = \frac{1}{g_{11}^*}\sqrt{g_{11}^*g_{11}\left(g_{11}^*g_{11}-1\right)};  \nonumber \\
h_{21}&=& \sqrt{\frac{\left(g_{11}^*g_{11}\right)^2-1}{g_{11}^*g_{11}}}; h_{22}=1.
\end{eqnarray}
\end{theorem}

\noindent
{\em Proof:}
Let the frequency $\omega$ be given.  In the sequel, we will not show the dependence on $j\omega$ for all transfer functions.   We first show that 
\begin{equation}
  \label{cost_bound}
g_{12}^*g_{12}+h_{12}^*h_{12} \geq g_{11}^*g_{11} - 1
\end{equation}
for all transfer function matrices (\ref{Gbar}), (\ref{GH}) satisfying the physical realizability condition (\ref{PR1}), the phase-insensitivity condition (\ref{PI}) and with the given amplifier gain $g_{11}$. Indeed, it follows by expanding out (\ref{PR1}) that the following equations are satisfied:
\begin{eqnarray}
\label{A}
  g_{11}^*h_{12}+g_{21}^*h_{22} &=& h_{11}g_{12}^*+h_{21}g_{22}^*; \\
\label{B}
  g_{12}^*h_{12}+g_{22}^*h_{22} &=& h_{12}g_{12}^*+h_{22}g_{22}^*; \\
  \label{C}
  g_{11}^*g_{11}+g_{21}^*g_{21} &=& h_{11}h_{11}^*+h_{21}h_{21}^*+1; \\
   \label{D}
  g_{11}^*g_{12}+g_{21}^*g_{22} &=& h_{11}h_{12}^*+h_{21}h_{22}^*; \\
    \label{E}
  g_{12}^*g_{12}+g_{22}^*g_{22} &=& h_{12}h_{12}^*+h_{22}h_{22}^*+1.                                                               
\end{eqnarray}
Now using the condition (\ref{PI}) and these equations, it is straightforward but tedious to verify that
\[
  h_{12}^*h_{12}-g_{12}^*g_{12} = g_{11}^*g_{11} - 1
\]
and hence
\begin{eqnarray*}
  g_{12}^*g_{12}+h_{12}^*h_{12}  &=& g_{11}^*g_{11} - 1+2g_{12}^*g_{12} \nonumber \\
                               &\geq& g_{11}^*g_{11} - 1.                                  
\end{eqnarray*}
That is, the equality (\ref{cost_bound}) is satisfied. Furthermore, equality holds when $g_{12} = 0$. Now it is straightforward to verify by substitution that if the transfer function elements $g_{12}$, $g_{21}$, $g_{22}$, $h_{11}$, $h_{12}$, $h_{22}$ are defined as in (\ref{optimal}), then the conditions (\ref{PR1}) and (\ref{PI}) will be satisfied. This completes the proof of the theorem.
\hfill $\Box$

In order to construct a physically realizable quantum system corresponding to a phase-insensitive amplifier whose DC transfer function matrix is derived from the above theorem, we will use the following lemma which is referred to as the Shale decomposition. 

\begin{lemma}[\cite{SHA62}, see also \cite{LN04,BRA05,GJN10,GP1}.]
\label{L1}
Consider a $4 \times 4$ complex matrix $\bar G$ of the form (\ref{Gbar}), (\ref{GH}) satisfying the physical realizability condition (\ref{PR1}). Then there exists a real diagonal matrix $R =  \left[\begin{array}{cc}r_1 & 0 \\0 & r_2\end{array}\right]$  and $2 \times 2$ unitary matrices $S_1$  and $S_2$ such that
\begin{eqnarray}
  \label{shale}
  \lefteqn{  \bar G =} \nonumber \\
  &&\hspace{-.3cm}\left[\begin{array}{cc}S_1 & 0 \\0 & S_1^\#\end{array}\right]
\left[\begin{array}{cc}-\cosh(R)&  -\sinh(R) \\
-\sinh(R)  & - \cosh(R)
      \end{array}\right]
    \left[\begin{array}{cc}S_2 & 0 \\0 & S_2^\#\end{array}\right].\nonumber \\
\end{eqnarray}
\end{lemma}
\ \\
Note that the decomposition given in this lemma is constructive; e.g., see \cite{LN04,BRA05}.

This lemma shows that the problem of physically realizing the two channel DC gain transfer function matrix $ \bar G$ can be reduced to the problem of physically realizing each of the single channel transfer function matrices
$\bar G_1 = \left[\begin{array}{cc}-\cosh(r_1)&  -\sinh(r_1) \\ -\sinh(r_1)  & - \cosh(r_1)\end{array}\right]$, and
$\bar G_2 = \left[\begin{array}{cc}-\cosh(r_2)&  -\sinh(r_2) \\ -\sinh(r_2)  & - \cosh(r_2)\end{array}\right]$. Then, the unitary transfer matrices $\left[\begin{array}{cc}S_1 & 0 \\0 & S_1^\#\end{array}\right]$, and $\left[\begin{array}{cc}S_2 & 0 \\0 & S_2^\#\end{array}\right]$ can be physically implemented using beamsplitters; e.g., see \cite{RZBB94,GP1}. Indeed, since $S_1$ and $S_2$ are both $2 \times 2$ matrices, it follows that each of these can be implemented by a single beamsplitter. For example, as in \cite{RZBB94} (with the addition of phase shifters on the input and output channels), we can write the input-output relations of a beamsplitter in the form
\[
\left[\begin{array}{c} y_1\\y_2\end{array}\right] = \mathcal{R} \left[\begin{array}{c} u_1\\u_2\end{array}\right]
\]
where $\mathcal{R}$ is a unitary matrix of the form
\begin{equation}
\label{mathcalR}
\mathcal{R} = \left[\begin{array}{cc}e^{j\phi_1}\sin(\theta) & e^{j(\phi_1+\phi_3)}\cos(\theta)\\
e^{j\phi_2}\cos(\theta) & -e^{j(\phi_2+\phi_3)}\sin(\theta)
\end{array}\right]
\end{equation}
and $\phi_1$, $\phi_2$, $\phi_3$  and $\theta$ are parameters of the beamsplitter. Furthermore, it is straightforward to verify that any  $2 \times 2$ unitary matrix $S$ can be represented as a matrix of the form (\ref{mathcalR}). 

To realize a single channel DC transfer function matrix
\begin{equation}
  \label{Gr}
  \bar G_r = \left[\begin{array}{cc}-\cosh(r)&  -\sinh(r) \\ -\sinh(r)  & - \cosh(r)\end{array}\right],
\end{equation}
  we consider a single channel dynamic squeezer following the approach of \cite{PET10Ca}; see also \cite{GJN10}. 

  An optical  cavity  consists of a number of  mirrors, one of which is partially reflective; e.g., see \cite{BR04,GZ00}. If we  include a nonlinear optical element inside such a cavity, an  optical squeezer can be obtained. By using suitable linearizations and approximations, such an optical squeezer can be
 described by
a  quantum stochastic differential equation as follows:
\begin{eqnarray}
\label{single_cavity}
da &=& -\frac{\kappa}{2}a dt -\chi a^* dt - \sqrt{\kappa} du;
  \nonumber \\
dy &=& \sqrt{\kappa}a dt + du,
\end{eqnarray}
where $\kappa > 0$, $\chi$ is a complex number associated with the strength of the nonlinear effect and $a$ is a single
annihilation operator associated with the cavity mode; e.g., see \cite{BR04,GZ00}. This leads to a linear quantum system 
of the form
(\ref{sys}) as follows:
\begin{eqnarray}
\label{example_qsde}
\left[\begin{array}{l} d a(t)\\d a(t)^*\end{array}\right] &=& 
\left[\begin{array}{ll}
-\frac{\kappa}{2}& -\chi \\ -\chi^* & -\frac{\kappa}{2}
\end{array}\right]
\left[\begin{array}{l}  a(t)\\ a(t)^*\end{array}\right]dt 
\nonumber \\&&
 - \sqrt{\kappa}
\left[\begin{array}{l}  du
\\ du^{*} \end{array}\right];  \nonumber \\
\left[\begin{array}{l} dy
\\ dy^{*} \end{array}\right] &=& 
\sqrt{\kappa}\left[\begin{array}{l}  a(t)\\ a(t)^*\end{array}\right]dt +
\left[\begin{array}{l}  du
\\ du^{*} \end{array}\right].
\end{eqnarray}
Note that it is straightforward to verify that this system is stable if and only if $\kappa^2 > 4\chi \chi^*$. 

As shown in \cite{PET10Ca}, these QSDEs are physically realizable with the corresponding $(S,L,\mathcal{H})$ parameters (e.g., see \cite{GJ09} for a discussion of $(S,L,\mathcal{H})$ parameters) given by
\[
S=I; N = \left[\begin{array}{cc} \sqrt{\kappa} & 0 \\ 0 & \sqrt{\kappa} \end{array}\right];  M = \left[\begin{array}{cc} 0 & -j \chi \\ j\chi^* & 0 \end{array}\right]. 
\]

\begin{figure}[htbp]
 \begin{center}
\includegraphics[width=8cm]{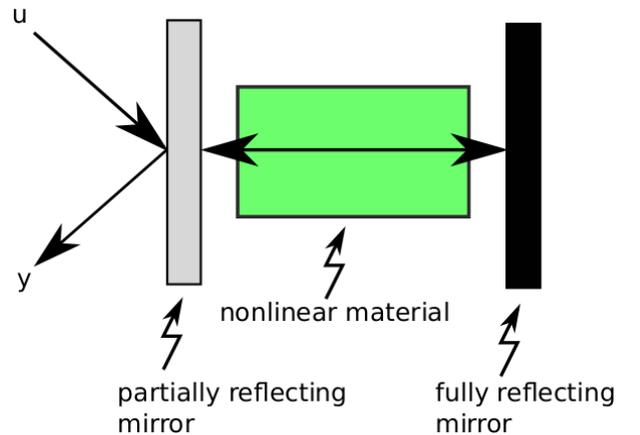}
 \end{center}
 \caption{Optical squeezer.}
 \label{F2}
\end{figure}

A diagram of a dynamic optical squeezer is shown in Figure \ref{F2}. 

Now, we choose the parameters $\kappa$ and $\chi$ to be of the form $\kappa = \epsilon \bar \kappa$ and $\chi = \epsilon \bar \chi$, where $\bar \kappa > 0$, $\bar \chi$ is chosen to be real and $\epsilon > 0$ is a parameter which will determine the amplifier bandwidth. Introducing the change of variables
$ \left[\begin{array}{l} \tilde a(t)\\  \tilde a(t)^*\end{array}\right] = \epsilon^{-\frac{1}{2}}\left[\begin{array}{l}  a(t)\\ a(t)^*\end{array}\right]$, the QSDEs (\ref{example_qsde}) reduce to 
\begin{eqnarray}
\label{example_qsde1}
\left[\begin{array}{l} d \tilde a(t)\\d \tilde a(t)^*\end{array}\right] &=& 
\frac{1}{\epsilon} \left[\begin{array}{ll}
-\frac{\bar \kappa}{2}& -\bar \chi \\ -\bar \chi & -\frac{\bar \kappa}{2}
\end{array}\right]
\left[\begin{array}{l}  \tilde a(t)\\ \tilde a(t)^*\end{array}\right]dt 
\nonumber \\&&
 - \frac{\sqrt{\bar \kappa}}{\epsilon}
\left[\begin{array}{l}  du
\\ du^{*} \end{array}\right];  \nonumber \\
\left[\begin{array}{l} dy
\\ dy^{*} \end{array}\right] &=& 
\sqrt{\bar \kappa}\left[\begin{array}{l}  \tilde a(t)\\ \tilde a(t)^*\end{array}\right]dt +
\left[\begin{array}{l}  du
\\ du^{*} \end{array}\right].
\end{eqnarray}

The transfer function matrix of this system at DC is given by
\begin{eqnarray*}
  G(0) &=& I+\left[\begin{array}{ll}
                   -\frac{2 \bar \kappa^2}{\bar \kappa^2 - 4 \bar \chi^2}& \frac{4\bar \kappa \bar \chi}{\bar \kappa^2 - 4 \bar \chi^2} \\
                    \frac{4\bar \kappa \bar \chi}{\bar \kappa^2 - 4 \bar \chi^2} & -\frac{2 \bar \kappa^2}{\bar \kappa^2 - 4 \bar \chi^2}
                   \end{array}\right]\nonumber \\
       &=&\left[\begin{array}{ll}-\frac{1+\alpha^2}{1-\alpha^2} & \frac{2 \alpha}{1-\alpha^2}\\
                   \frac{2 \alpha}{1-\alpha^2} & -\frac{1+\alpha^2}{1-\alpha^2}\end{array}\right]
\end{eqnarray*}
where $\alpha =\frac{2 \bar \chi}{\bar \kappa}= \frac{2  \chi}{ \kappa}$. In order for the system to be stable, we require $\alpha^2 < 1$. In order to construct a physically realizable quantum system with DC transfer function matrix $\bar G_r$ defined in (\ref{Gr}), we equate $G(0)$ with $\bar G_r$. That is,
\[
\left[\begin{array}{ll}-\frac{1+\alpha^2}{1-\alpha^2} & \frac{2 \alpha}{1-\alpha^2}\\
                   \frac{2 \alpha}{1-\alpha^2} & -\frac{1+\alpha^2}{1-\alpha^2}\end{array}\right] =  \left[\begin{array}{cc}-\cosh(r)&  -\sinh(r) \\ -\sinh(r)  & - \cosh(r)\end{array}\right].
  \]
  This is equivalent to the equations
  \begin{eqnarray*}
    \cosh(r) &=& \frac{1+\alpha^2}{1-\alpha^2}; \\
    \sinh(r) &=&  -\frac{2 \alpha}{1-\alpha^2}
  \end{eqnarray*}
  To see that these equations are consistent, we calculate
  \[
\cosh^2(r)- \sinh^2(r) = \frac{1+2\alpha^2+\alpha^4}{(1-\alpha^2)^2}- \frac{4 \alpha^2}{(1-\alpha^2)^2} = 1
\]
as required.

Now given $r$, we construct the corresponding value of $\alpha$ satisfying $\alpha^2 < 1$ such that $ \sinh(r) =  -\frac{2 \alpha}{1-\alpha^2}$. This is equivalent to the equation
\[
\alpha^2 - \frac{2\alpha}{\sinh(r)} - 1 =0.
  \]
  This equation has two solutions:
  \[
    \alpha = \frac{1+\cosh(r)}{\sinh(r)}= \frac{1}{\tanh(\frac{r}{2})}
  \]
  and
  \[
    \alpha = \frac{1-\cosh(r)}{\sinh(r)}= -\tanh(\frac{r}{2}).
  \]
  However, $\tanh(\frac{r}{2}) \in (-1,1)$ and hence only the solution
  \begin{equation}
    \label{design1}
\alpha =-\tanh(\frac{r}{2})
\end{equation}
satisfies the condition $\alpha^2 < 1$.
    
The above discussion leads to the following result.

\begin{lemma}
  \label{L2}
  Given any matrix $G_r$ of the form (\ref{Gr}), there exists a physically realizable quantum system of the form (\ref{example_qsde1}) corresponding to a stable single channel dynamic squeezer such that its transfer function matrix $G(s)$ satisfies
\[
  G(0) = G_r.
\]
Here, the ratio $\alpha =\frac{2 \bar \chi}{\bar \kappa}$ satisfying $\alpha^2 < 1$ is uniquely determined by the design equation 
(\ref{design1}) and the parameter $\epsilon > 0$ can be chosen to achieve any desired bandwidth.
\end{lemma}

We now combine Theorem \ref{T1} with Lemmas \ref{L1} and \ref{L2} to obtain the following theorem which is our main result.

\begin{theorem}
  \label{T2}
  Given any desired quantum phase-insensitive amplifier DC gain $g_{11}$, there exists a corresponding physically realizable linear quantum system of the form (\ref{sys}) which achieves this DC gain and introduces the minimal amount of DC quantum noise defined by (\ref{noise_bound}). Furthermore, this transfer function matrix satisfies the DC phase-insensitivity condition (\ref{PI}).  In addition, the parameters in this linear quantum system can be chosen to achieve a specified bandwidth over which the above conditions will hold approximately. Finally, this system can be constructed from two beamsplitters and two stable dynamic squeezers of the form (\ref{example_qsde}).
\end{theorem}

\section{Illustrative Example}
We now apply the method of this paper to synthesize a phase-insensitive quantum amplifier with a DC gain of $g_{11} = 2$ (6dB), a bandwidth of $2 \times 10^6$ radians/s  and with the minimum added noise. Indeed, with $g_{11} = 2$, the formulas (\ref{optimal}) give
\[
G= \left[\begin{array}{cc}2 & 0 \\
                    \frac{\sqrt{3}}{2} & \sqrt{5} \end{array}\right]; \quad H = \left[\begin{array}{cc}0 & \sqrt{3} \\
                    \frac{\sqrt{15}}{2} & 1 \end{array}\right].
              \]
              We then apply Lemma \ref{L1}. This leads to the equations (\ref{shale}) where
              \[
                R= \left[\begin{array}{cc}    1.6139 &        0 \\
         0  &  -1.1327     \end{array}\right].
   \]
   Also, we have
   \[
     S_1 = \left[\begin{array}{cc} 0.5240 &   0.8517 \\
    0.8517 &  -0.5240\end{array}\right]
\]
and
\[
     S_2   = \left[\begin{array}{cc} -0.6840 &   -0.7295\\
                                -0.7295  &  0.6840\end{array}\right].                            
                          \]

\begin{figure}[htbp]
 \begin{center}
\includegraphics[width=8cm]{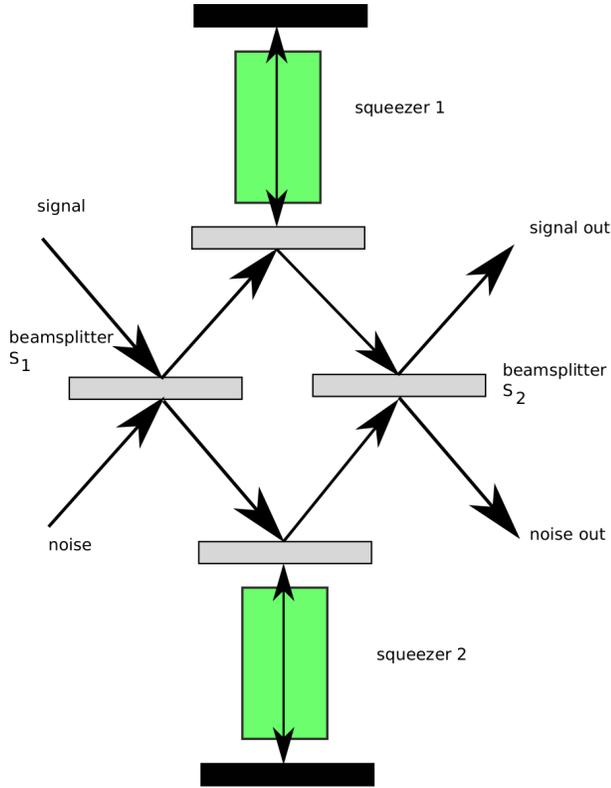}
 \end{center}
 \caption{Proposed realization of the phase-insensitive quantum amplifier.}
 \label{F3}
\end{figure}

                          Now,  we observe that the matrix $S_1$ is a matrix of the form (\ref{mathcalR}) with parameters, $\theta_1 = 0.5515$ radians and  $\phi_1= 0$, $\phi_2=0 $, $\phi_3=0$. Similarly, the $S_2$ is a matrix of the form (\ref{mathcalR}) with parameters, $\theta_1 = -0.7532$ radians and  $\phi_1= 0$, $\phi_2=\pi $, $\phi_3=\pi$. These parameter values define the beamsplitters representing the matrices $S_1$ and $S_2$ respectively. Also, the matrix $R$ defines the parameters $\alpha_1= -0.6679$ and $\alpha_2 = 0.5127$ according to the formula (\ref{design1}). These parameters are then used to define the parameters for the two squeezers. First we choose the parameter $\epsilon = 2\pi 10^6$ radians/s  to achieve the specified bandwidth. Then, we choose the parameters $\kappa_1 = 2\pi*10^6$ radians/s, $\chi_1 = \frac{\alpha_1 \kappa_1}{2} = -2.0983\times 10^6$ radians/s for the first squeezer, and the parameters  $\kappa_2 = 2\pi*10^6$ radians/s, $\chi_2 = \frac{\alpha_2 \kappa_2}{2} = 1.6106\times 10^6$ radians/s for the second squeezer. The implementation of the phase-insensitive amplifier is  as shown in Figure \ref{F3}.

\begin{figure}[htbp]
 \begin{center}
\includegraphics[width=8cm]{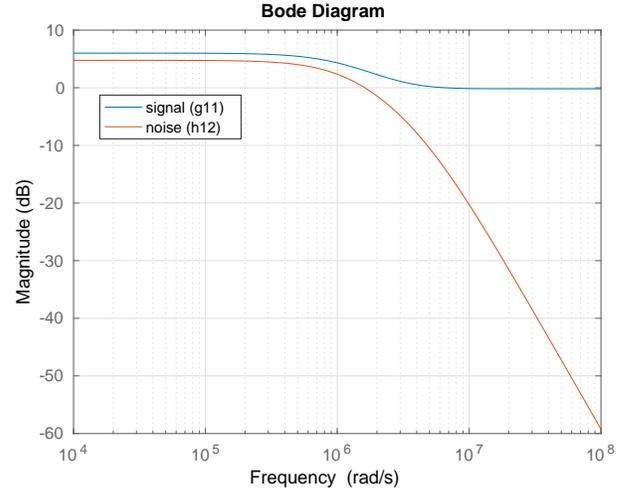}
 \end{center}
 \caption{Magnitude Bode plots of the phase-insensitive quantum amplifier signal and noise gain.}
 \label{F4}
\end{figure}

We now calculate the transfer function matrix of this proposed phase-insensitive quantum amplifier. Let $\tilde G_1(s)$ be the transfer function of the first squeezer, defined by state equations of the form (\ref{example_qsde}) and let $\tilde G_2(s)$ be the transfer function of the second squeezer, also defined by state equations of the form (\ref{example_qsde}). Then it is straightforward to verify that the transfer function matrix of the overall phase-insensitive quantum amplifier system is given by
\begin{eqnarray*}
  \lefteqn{\bar G(s) =} \nonumber \\
  &&\left[\begin{array}{cc}S_1 & 0 \\0 & S_1^\#\end{array}\right]
  \left[\begin{array}{cccc}1 & 0 & 0 & 0 \\
          0 & 0 & 1 & 0 \\
          0 & 1 & 0 & 0 \\
          0 & 0 & 0 & 1 \\ 
        \end{array}\right]
       \nonumber \\
            && \times \left[\begin{array}{cc}\tilde G_1(s) & 0 \\0 & \tilde G_2(s) \end{array}\right]
    \left[\begin{array}{cccc}1 & 0 & 0 & 0 \\
          0 & 0 & 1 & 0 \\
          0 & 1 & 0 & 0 \\
          0 & 0 & 0 & 1 \\
            \end{array}\right]
          \left[\begin{array}{cc}S_2 & 0 \\0 & S_2^\#\end{array}\right]. 
\end{eqnarray*}
We construct this transfer function matrix for this example and then plot the magnitude Bode plot of the $(1,1)$ block of $\bar G(s)$ as shown in Figure \ref{F4}. This is the transfer function from the signal input to the signal output $g_{11}(s)$. This plot also shows the magnitude Bode plot of the $(1,4)$ block of $\bar G(s)$. This is the transfer function from the quadrature noise input to the signal output $h_{12}(s)$. This plot shows that at DC, the amplifier gives 6 dB of gain but there is a noise signal which is of a magnitude given by the formula (\ref{noise_bound}).

\bibliography{/home/irp/Bibliog/irpnew}

\begin{thebibliography}{10}
\providecommand{\url}[1]{#1}
\csname url@samestyle\endcsname
\providecommand{\newblock}{\relax}
\providecommand{\bibinfo}[2]{#2}
\providecommand{\BIBentrySTDinterwordspacing}{\spaceskip=0pt\relax}
\providecommand{\BIBentryALTinterwordstretchfactor}{4}
\providecommand{\BIBentryALTinterwordspacing}{\spaceskip=\fontdimen2\font plus
\BIBentryALTinterwordstretchfactor\fontdimen3\font minus
  \fontdimen4\font\relax}
\providecommand{\BIBforeignlanguage}[2]{{%
\expandafter\ifx\csname l@#1\endcsname\relax
\typeout{** WARNING: IEEEtran.bst: No hyphenation pattern has been}%
\typeout{** loaded for the language `#1'. Using the pattern for}%
\typeout{** the default language instead.}%
\else
\language=\csname l@#1\endcsname
\fi
#2}}
\providecommand{\BIBdecl}{\relax}
\BIBdecl

\bibitem{GJN10}
J.~E. Gough, M.~R. James, and H.~I. Nurdin, ``Squeezing components in linear
  quantum feedback networks,'' \emph{Physical Review A}, vol.~81, p. 023804,
  2010.

\bibitem{ZJ11}
G.~Zhang and M.~James, ``Direct and indirect couplings in coherent feedback
  control of linear quantum systems,'' \emph{IEEE Transactions on Automatic
  Control}, vol.~56, no.~7, pp. 1535--1550, 2011.

\bibitem{ShP5}
A.~J. Shaiju and I.~R. Petersen, ``A frequency domain condition for the
  physical realizability of linear quantum systems,'' \emph{IEEE Transactions
  on Automatic Control}, vol.~57, no.~8, pp. 2033 -- 2044, 2012.

\bibitem{PET10B}
I.~R. Petersen, ``Quantum linear systems theory,'' \emph{Open Automation and
  Control Systems Journal}, vol.~8, pp. 67--93, 2016.

\bibitem{NY17}
H.~I. Nurdin and N.~Yamamoto, \emph{Linear Dynamical Quantum Systems: Analysis,
  Synthesis, and Control}.\hskip 1em plus 0.5em minus 0.4em\relax Berlin:
  Springer, 2017.

\bibitem{CAV82}
C.~M. Caves, ``Quantum limits on noise in linear amplifiers,'' \emph{Physical
  Review D}, vol.~26, no.~8, pp. 1817--1839, 1982.

\bibitem{HM62}
H.~Haus and J.~Mullen, ``Quantum noise in linear amplifiers,'' \emph{Physical
  Review}, vol. 128, pp. 2407--2413, 1962.

\bibitem{BSM+7}
N.~Bergeal, F.~Schackert, M.~Metcalfe, R.~Vijay, V.~Manucharyan, L.~Frunzio,
  D.~Prober, R.~Schoelkopf, S.~Girvin, and M.~Devoret, ``Phase-preserving
  amplification near the quantum limit with a josephson ring modulator,''
  \emph{Nature}, vol. 465, pp. 64--68, 2010.

\bibitem{CDGMS10}
A.~A. Clerk, M.~Devoret, S.~Givin, F.~Marquardt, and R.~Schoelkopf,
  ``Introduction to quantum noise, measurement, and amplification,''
  \emph{Reviews of Modern Physics}, vol.~82, p. 1155, 2010.

\bibitem{CWA+5}
H.~M. Chrzanowski, N.~Walk, S.~M. Assad, J.~Janousek, S.~Hosseini, T.~C. Ralph,
  T.~Symul, and P.~K. Lam, ``Measurement-based noiseless linear amplification
  for quantum communication,'' \emph{Nature Photonics}, vol.~8, no.~4, pp.
  333--338, 2014.

\bibitem{MC14}
A.~Metelmann and A.~A. Clerk, ``Quantum-limited amplification via reservoir
  engineering,'' \emph{Phys. Rev. Lett.}, vol. 112, p. 133904, Apr 2014.

\bibitem{HKL+3}
F.~Hudelist, J.~Kong, C.~Liu, J.~Jing, Z.~Ou, and W.~Zhang, ``Quantum metrology
  with parametric amplifier-based photon correlation interferometers,''
  \emph{Nature Communications}, vol.~5, p. 3049, 2014.

\bibitem{YAM16}
N.~Yamamoto, ``Quantum feedback amplification,'' \emph{Physical Review
  Applied}, vol.~5, p. 044012, 2016.

\bibitem{YMFF11}
J.~ichi Yoshikawa, Y.~Miwa, R.~Filip, and A.~Furusawa, ``Demonstration of a
  reversible phase-insensitive optical amplifier,'' \emph{Physical Review A},
  vol.~83, p. 052307, May 2011.

\bibitem{GGY08}
J.~Gough, R.~Gohm, and M.~Yanagisawa, ``Linear quantum feedback networks,''
  \emph{Physical Review A}, vol.~78, p. 062104, 2008.

\bibitem{KhP1}
A.~Khodaparastsichani and I.~R. Petersen, ``A modified frequency domain
  condition for the physical realizability of linear quantum stochastic
  systems,'' \emph{IEEE Transactions on Automatic Control}, 2018, to appear,
  accepted 25 June 2017.

\bibitem{PET09A}
I.~R. Petersen, ``Singular perturbation approximations for a class of linear
  quantum systems,'' \emph{IEEE Transactions on Automatic Control}, vol.~58,
  no.~1, pp. 193--198, 2013, arXiv:1107.5605.

\bibitem{VuP3a}
S.~L. Vuglar and I.~R. Petersen, ``Singular perturbation approximations for
  general linear quantum systems,'' in \emph{Proceedings of the 2012 Australian
  Control Conference}, Sydney, Australia, November 2012, arXiv:1208.6155.

\bibitem{DLMGNL15}
A.~Dutt, K.~Luke, S.~Manipatruni, A.~L. Gaeta, P.~Nussenzveig, and M.~Lipson,
  ``On-chip optical squeezing,'' \emph{Physics Review Applied}, vol.~3, p.
  044005, Apr 2015.

\bibitem{BHJ07}
L.~Bouten, R.~{van~Handel}, and M.~James, ``An introduction to quantum
  filtering,'' \emph{SIAM J. Control and Optimization}, vol.~46, no.~6, pp.
  2199--2241, 2007.

\bibitem{PAR92}
K.~Parthasarathy, \emph{An Introduction to Quantum Stochastic Calculus}.\hskip
  1em plus 0.5em minus 0.4em\relax Berlin: Birkhauser, 1992.

\bibitem{HP84}
R.~Hudson and K.~Parthasarathy, ``Quantum \mbox{I}to's formula and stochastic
  evolution,'' \emph{Communications in Mathematical Physics}, vol.~93, pp.
  301--323, 1984.

\bibitem{VPB92}
V.~Belavkin, ``Quantum continual measurements and a posteriori collapse on
  {CCR},'' \emph{Commun. Math. Phys.}, vol. 146, pp. 611--635, 1992.

\bibitem{JNP1}
M.~R. James, H.~I. Nurdin, and I.~R. Petersen, ``${H}^\infty$ control of linear
  quantum stochastic systems,'' \emph{IEEE Transactions on Automatic Control},
  vol.~53, no.~8, pp. 1787--1803, 2008.

\bibitem{GZ00}
C.~Gardiner and P.~Zoller, \emph{Quantum Noise}.\hskip 1em plus 0.5em minus
  0.4em\relax Berlin: Springer, 2000.

\bibitem{GJ09}
J.~Gough and M.~R. James, ``The series product and its application to quantum
  feedforward and feedback networks,'' \emph{IEEE Transactions on Automatic
  Control}, vol.~54, no.~11, pp. 2530--2544, 2009.

\bibitem{EB05}
S.~C. Edwards and V.~P. Belavkin, ``Optimal quantum feedback control via
  quantum dynamic programming,'' University of Nottingham, quant-ph/0506018,
  2005.

\bibitem{NJP1}
H.~I. Nurdin, M.~R. James, and I.~R. Petersen, ``Coherent quantum {LQG}
  control,'' \emph{Automatica}, vol.~45, no.~8, pp. 1837--1846, 2009.

\bibitem{NJD09}
H.~I. Nurdin, M.~R. James, and A.~C. Doherty, ``Network synthesis of linear
  dynamical quantum stochastic systems,'' \emph{SIAM Journal on Control and
  Optimization}, vol.~48, no.~4, pp. 2686--2718, 2009.

\bibitem{PET08A}
I.~R. Petersen, ``Cascade cavity realization for a class of complex transfer
  functions arising in coherent quantum feedback control,'' \emph{Automatica},
  vol.~47, no.~8, pp. 1757--1763, 2011.

\bibitem{NUR10}
H.~Nurdin, ``Synthesis of linear quantum stochastic systems via quantum
  feedback networks,'' \emph{IEEE Transactions on Automatic Control}, vol.~55,
  no.~4, pp. 1008 --1013, April 2010.

\bibitem{NUR10A}
------, ``On synthesis of linear quantum stochastic systems by pure
  cascading,'' \emph{IEEE Transactions on Automatic Control}, vol.~55, no.~10,
  pp. 2439 --2444, October 2010.

\bibitem{NGP1}
H.~I. Nurdin, S.~Grivopoulos, and I.~R. Petersen, ``The transfer function of
  generic linear quantum stochastic systems has a pure cascade realization,''
  \emph{Automatica}, vol.~69, p. 324–333, July 2016.

\bibitem{GP1}
S.~Grivopoulos and I.~R. Petersen, ``Linear quantum system transfer function
  realization using static networks for i/o processing and feedback,''
  \emph{SIAM Journal on Control and Optimization}, vol.~55, no.~5, pp.
  3349--3369, 2017.

\bibitem{SHA62}
D.~Shale, ``Linear symmetries of free boson fields,'' \emph{Transactions of the
  American Mathematical Society}, vol. 103, pp. 149--167, 1962.

\bibitem{LN04}
U.~Leonhardt and A.~Neumaier, ``Explicit effective hamiltonians for general
  linear quantum-optical networks,'' \emph{Journal of Optics B: Quantum and
  Semiclassical Optics}, vol.~6, pp. L1--L4, 2004.

\bibitem{BRA05}
S.~L. Braunstein, ``Squeezing as an irreducible resource,'' \emph{PHYSICAL
  REVIEW A}, vol.~71, p. 055801, 2005.

\bibitem{RZBB94}
M.~Reck, A.~Zeilinger, H.~Bernstein, and P.~Bertani, ``Experimental realization
  of any discrete unitary operator,'' \emph{Physical Review Letters}, vol.~73,
  no.~1, pp. 58--61, 1994.

\bibitem{PET10Ca}
I.~R. Petersen, ``Realization of single mode quantum linear systems using
  static and dynamic squeezers,'' in \emph{Proceedings of the 8th Asian Control
  Conference}, Kaohsiung, Taiwan, May 2011.

\bibitem{BR04}
H.~Bachor and T.~Ralph, \emph{A Guide to Experiments in Quantum Optics},
  2nd~ed.\hskip 1em plus 0.5em minus 0.4em\relax Weinheim, Germany: Wiley-VCH,
  2004.

\end{thebibliography}
\bibliographystyle{IEEEtran}

\end{document}